# Detecting Stripped Stars While Searching for Quiescent Black Holes

Julia Bodensteiner[1]
and Marianne Heida[1]
Michael Abdul-Masih[1]
Dietrich Baade[1]
Gareth Banyard[2]
Dominic M. Bowman[2]
Matthias Fabry[2]
Abigail Frost[2]
Laurent Mahy[3]
Pablo Marchant[2]
Antoine Mérand[1]
Maddalena Reggiani[2]
Thomas Rivinius[1]
Hugues Sana[2]
Fernando Selman[1]
Tomer Shenar[4]

[1] ESO
[2] Institute of Astronomy, KU Leuven, Belgium
[3] Royal Observatory of Belgium, Brussels, Belgium
[4] Anton Pannekoek Institute for Astronomy, University of Amsterdam, the Netherlands

While the number of stellar-mass black holes detected in X-rays or as gravitational wave sources is steadily increasing, the known population remains orders of magnitude smaller than predicted by stellar evolution theory. A significant fraction of stellar-mass black holes is expected to hide in X-ray-quiet binaries where they are paired with a "normal" star. Although a handful of such quiescent black hole candidates have been proposed, the majority have been challenged by follow-up investigations. A confusion that emerged recently concerns binary systems that appear to contain a normal B-type star with an unseen companion, believed to be a black hole. On closer inspection, some of these seemingly normal B-type stars instead turn out to be stars stripped of most of their mass through an interaction with their binary companion, which in at least two cases is a rapidly rotating star rather than a compact object. These contaminants in the search for quiescent black holes are themselves extremely interesting objects as they represent a rare phase of binary evolution, and should be given special attention when searching for binaries hosting black holes in large spectroscopic studies.

## Black holes as tracers of massive-star evolution

Massive stars play an important role in the Universe. They are sources of energetic radiation, forge heavy chemical elements from the nuclear reactions occurring in their cores, and provide dynamical and chemical feedback via their strong stellar winds. Stars massive enough to undergo core collapse may end their lives in a supernova explosion, releasing the freshly formed elements and possibly leaving behind a neutron star or a stellar-mass black hole. They can also collapse directly into a black hole, or explode in a pair-instability supernova without leaving a remnant behind (see, for example, Heger et al., 2003; Sukhbold et al., 2016). The specific factors that dictate which stars will reach a particular fate are not yet completely understood but they have implications for the type of supernova that will occur, the properties of the remnant they leave behind, and the enrichment of the interstellar medium (see, for example, Farmer et al., 2021; Laplace et al., 2021). Apart from the supernovae themselves, the distributions of black hole masses, spins and natal kicks provide important clues about the final stages of massive stars. Studying these ideally requires a large and unbiased sample of stellar-mass black holes, which is difficult to build up observationally.

Given that black holes are intrinsically dark objects, detecting them on their own is almost impossible. Currently, black holes are discovered mainly in two ways: as X-ray binaries, when they accrete material from a companion star, or as gravitational wave sources, when they merge with another black hole or neutron star. The black holes found in Galactic X-ray binaries tend to have lower masses (~ 5–20 $M_\odot$, Corral-Santana et al., 2016), while the black holes found by the Laser Interferometer Gravitational-Wave Observatory and its European counterpart (LIGO/Virgo) span a much larger mass range (3–100 $M_\odot$; see Figure 1 and, for example, Fishbach & Holz, 2017). The two populations are subject to different observational and evolutionary selection biases (for example, Tauris & van den Heuvel, 2006; Mandel & Farmer, 2018).

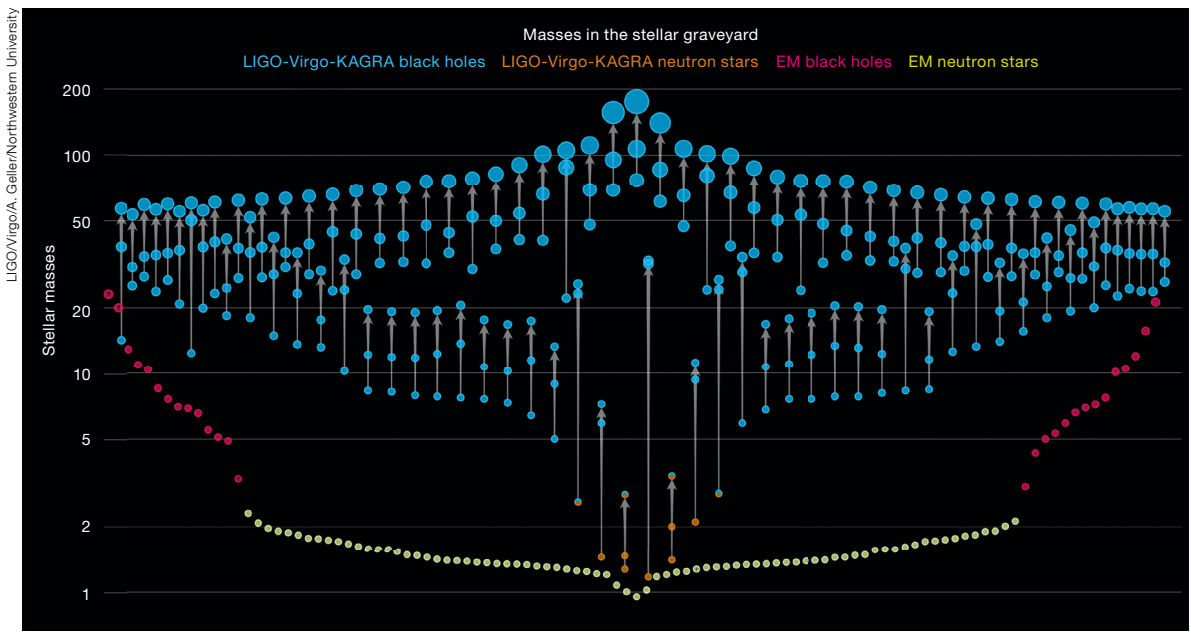

Figure 1. Overview of black hole and neutron star masses detected using different techniques including gravitational wave detections by LIGO/Virgo/KAGRA (bright blue) and electromagnetic detections mainly in X-ray binaries (dark red). While the gravitational wave detections span the mass range ~ 3–100 $M_\odot$ for single black holes, those detected in X-ray binaries only go up to ~ 21 $M_\odot$.

But these accreting and merging black holes are only the tip of the iceberg. Since most of the massive stars that have formed throughout the history of the Universe end their lives as black holes, and there is basically no mechanism to destroy them, many more of them should be out there. Population synthesis calculations predict a population of order $10^7$ black holes in the Milky Way alone (van den Heuvel, 1992), of which only about 100 (confirmed and candidate) have been discovered in X-ray binaries. The vast majority of the remaining black holes are not accreting enough material to show up in X-ray observations and are therefore called quiescent black holes. There are several ways to potentially detect these: microlensing is a promising technique (Paczynski, 1986; Wyrzykowski & Mandel, 2020), as are astrometric detections using Gaia data (Breivik, Chatterjee & Larson, 2017; Janssens et al., 2021). In addition, it has been suggested that periodic variability of the orbital periods of several close binaries is due to a quiescent black hole as a tertiary component (Qian, Liao & Fernández Lajús, 2008; Liao & Qian, 2010; Er-Gang et al., 2019; Wang & Zhu, 2021). However, most of the candidates identified to date were found in spectroscopic studies by detecting the motion of a star orbiting a dark, unidentified object.

## The spectroscopic search for quiescent black holes

Binary systems offer unique possibilities for determining the masses of stars and compact objects, for example using phase-resolved spectroscopic observations. If both components of a binary are visible in the spectra and their lines show radial velocity (RV) variations in antiphase with each other (also referred to as reflex motion), their mass ratio is simply the inverse of the ratio of their radial velocity semi-amplitudes. If, in addition to the mass ratio, the mass of one of the components of the binary is known, the mass of the second object automatically follows. A second way to obtain information on the masses of binary stars that is also useful for systems where only one object is visible in the spectra, is via the so-called binary mass function. This only requires knowledge of the orbital period and the radial velocity semi-amplitude of one of the binary components, and provides a strict lower limit on the mass of the other component (see, for example, Casares & Jonker, 2014). To obtain the actual mass of this second object, either the mass ratio of the binary or the mass of the first object has to be determined, as well as the orbital inclination.

When searching for X-ray-quiet stellar-mass black holes one should therefore look for stars with approximately sinusoidal radial velocity variations, indicating that they are part of a binary, but without a visible companion showing a reflex motion. The best candidates are early-type stars (which are generally more massive) with large radial velocity amplitudes and long orbital periods. These characteristics combined imply a high mass of the unseen companion. However, there are many possible reasons why this unseen companion cannot be discerned in the spectra other than its being a black hole. It could be a star that is not detected either because of low-quality data, or because it is faint compared to the initially discovered star. Another possibility is that the companion is rotating rapidly, causing its absorption lines to be broad and shallow.

Large surveys, both in train and planned, will provide extensive homogeneous spectroscopic datasets that enable scientists to mine for quiescent black holes (Gu et al., 2019; Yi, Sun & Gu, 2019). Surveys currently underway include those carried out with the Large Sky Area Multi-Object Fiber Spectroscopic Telescope (LAMOST; Cui et al., 2012) and the Sloan Digital Sky Survey (SDSS; York et al., 2000). The near future will see surveys using the 4-metre Multi-Object Spectrograph Telescope (4MOST) which will enter operation in 2023 at the VISTA telescope in Paranal (de Jong et al., 2012) and the WHT Enhanced Area Velocity Explorer (WEAVE; Dalton et al., 2012) at the William Herschel Telescope. While the detection principle of quiescent black holes is simple, one should be aware of the contaminants mentioned above. Their diversity is illustrated by the competing explanations put forward for several of the candidate quiescent black holes discovered to date. Here we provide a short overview of the current sample of candidate quiescent black holes and their alternative interpretations, and then focus on one particular class of contaminants that is extremely interesting in its own right, namely stripped B-type stars with a rapidly rotating companion star.

## Candidate quiescent black holes

Only a handful of candidate quiescent black holes have been reported to date. Three systems strongly suspected to contain a black hole with a relatively massive companion star are, MWC 656, a binary system of a black hole and a classical Be star (Casares et al., 2014); AS 386, a B[e] star with a ~ 7 $M_\odot$ dark companion (Khokhlov et al., 2018); and HD 96670, a ~ 22 $M_\odot$ O star in a close binary with a ~ 7 $M_\odot$ black hole and possibly a third component (Gomez & Grindlay, 2021). Both MWC 656 (Munar-Adrover et al., 2014) and HD 96670 (Grindlay et al., in preparation) were subsequently found to show faint X-ray emission with a hard power-law spectrum consistent with very low-level accretion onto a compact object; MWC 656 has also been detected in the radio, further strengthening its identification as a black hole accreting at a very low level (Dzib, Massi & Jaron, 2015). None of these three quiescent black holes has thus far been disputed. A recent addition to the list of black hole companions to massive stars is NGC 2004#115 (Lennon et al., 2021), a B-type star with a suggested 25 $M_\odot$ black hole companion. However, the presence of a massive compact object in this system hinges on the inferred orbital inclination of only 9 degrees, which has been challenged by El-Badry, Burdge & Mroz (2021).

There are six candidate quiescent black holes with low-mass stellar companions.

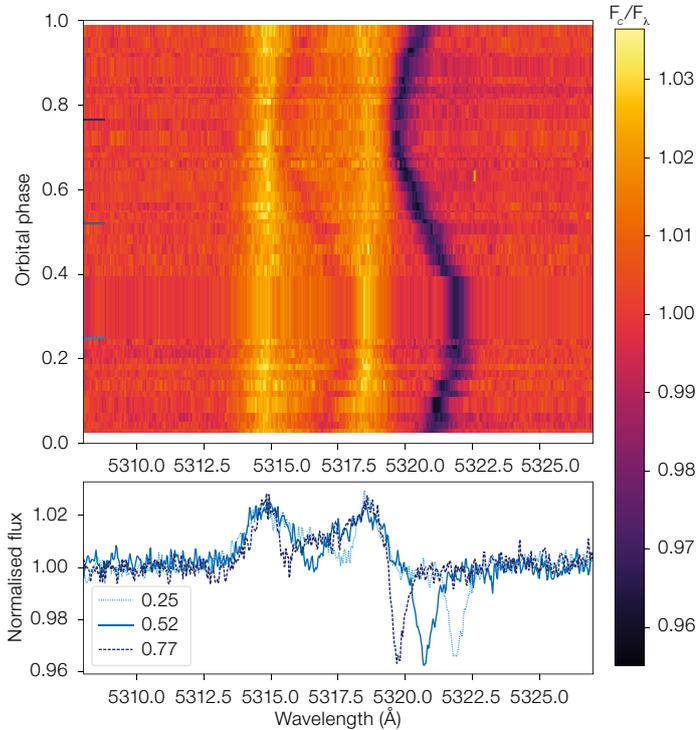

Figure 2. Phase-resolved spectra of HR 6819 showing the region around the Fe II line at 5316 Å, covering the orbital period of 40 days. The top panel shows the dynamical spectrum while the bottom panel shows three individual normalised spectra at phase 0.25, 0.52 and 0.77 as indicated in the legend. The double-peaked emission line (in bright), which seems to be stationary, is attributed to the Be star in the system, while the periodically shifting absorption lines (in dark) are attributed to the B-type star.

Three are located in the Galactic globular cluster NGC 3201, with minimum masses of ~ 4, ~ 4 and ~ 8 $M_\odot$, respectively (Giesers et al., 2018, 2019. Two systems, and V723 Mon and 2MASS J05215658+4359220, are red giants with a ~ 3 $M_\odot$ dark companion (Thompson et al., 2019; Jayasinghe et al., 2021; Masuda & Hirano, 2021). However, it has been pointed out that at least the latter system could also be a triple system with a giant tertiary star and an inner binary consisting of two main-sequence stars (van den Heuvel & Tauris, 2020; and see Thompson et al., 2020 for a response). The existence of a third, very similar system consisting of an evolved red giant with a 2–3 $M_\odot$ dark companion was proposed recently (Jayasinghe et al., 2022).

Finally, there are three systems, LB-1 (Liu et al., 2019), HR 6819 (Rivinius et al., 2020) and NGC 1850 BH1 (Saracino et al., 2021) that were recently reported as quiescent black holes accompanied by a B-type star. However, it was subsequently proposed that they are instead binary systems consisting of a stripped B-type star and another luminous star (Shenar et al., 2020b; Bodensteiner, et al., 2020; El-Badry & Quataert, 2021; El-Badry & Burdge, 2021; Frost et al., 2022). The nature of NGC 1850 BH1 is still highly debated: El-Badry & Burdge (2021) point out that, given the orbital parameters, the mass of the radial-velocity variable star must be much lower than that assumed by Saracino et al. 2021. This implies a lower mass for the unseen companion that moves it out of the black hole mass range. Stevance, Parsons & Eldridge (2021) add that theoretically the black hole scenario is very unlikely. We therefore focus on HR 6819 and LB-1 in the following.

## The spectroscopic signature of LB-1 and HR 6819

LB-1 and HR 6819 share a specific spectroscopic signature. They both show narrow absorption lines indicative of a B-type star moving on a period of tens of days (that we will refer to as the RV variable star), as well as broad emission lines that appear stationary (see Figure 2). The interpretation of these systems hinges on 1) the source of the emission features, which could originate in an accretion disc, a circumbinary disc, or a decretion disc around a rapidly rotating star — a so-called classical Be star (Rivinius, Carciofi & Martayan, 2013) — and 2) whether the emission features are indeed stationary or instead show a small reflex motion with respect to the B-type star. If they are moving in anti-phase, then the B-type star and the source of the emission features form a binary system. If the emission lines are truly stationary, the B-type star orbits an unseen object and the source of the emission lines is either an unrelated star associated only by chance superposition, or it is a third, outer component, making the system a triple system.

The detection of rotationally broadened photospheric absorption lines in the spectrum in combination with the emission lines is evidence for the presence of a classical Be star, and thus settles 1) above. However, these absorption lines are difficult to detect as they are very shallow. Settling 2) above is also challenging as the emission features are broad and affected by superimposed, moving absorption lines from the B-type star, making a small radial velocity amplitude hard to detect. This is additionally complicated by the fact that the two stars have similar temperatures and therefore the same set of spectral lines.

## Black holes or stripped stars in LB-1 and HR 6819?

In 2019 LB-1 made headlines as a binary system of a B-type star and a 70 $M_\odot$ black hole (Liu et al., 2019). They reported a small reflex motion of the H-alpha emission line but did not detect broad absorption lines. Interpreting the emission features as evidence for an accretion disc around a black hole and assuming a typical mass of ~ 5 $M_\odot$ for the B-type star then led to the large inferred black hole mass. As this exceeded predictions from stellar evolution theory, the finding triggered an intense debate. While theorists tried to reconcile theory and observations by adapting their black hole mass predictions (Groh et al., 2020; Eldridge et al., 2020; Safarzadeh, Ramirez-Ruiz & Kilpatrick, 2020; Belczynski et al., 2020), observers pointed out issues with the initial data analysis, in particular with the claim of reflex motion of the H-alpha line (Abdul-Masih et al., 2020; El-Badry & Quataert, 2020; Simón-Díaz et al., 2020; Irrgang et al., 2020). Based on new near-infrared data, Liu et al. (2020) refined their analysis and reduced the claimed mass of the unseen companion. Using additional optical spectra, Shenar et al. (2020b) eventually detected broadened absorption lines, showing that instead of a black hole, the second object in the system is a Be star, although Lennon et al. (2021) have challenged this interpretation based on Hubble Space Telescope spectroscopy.

Figure 3. Schematic comparison between the binary and the triple scenario proposed for HR 6819. The projected distances on the sky between the two visible stars, the stripped star and the Be star in the binary scenario, and the B-type giant and the Be star in the triple scenario, are indicated in blue (Figures and symbols are not to scale).

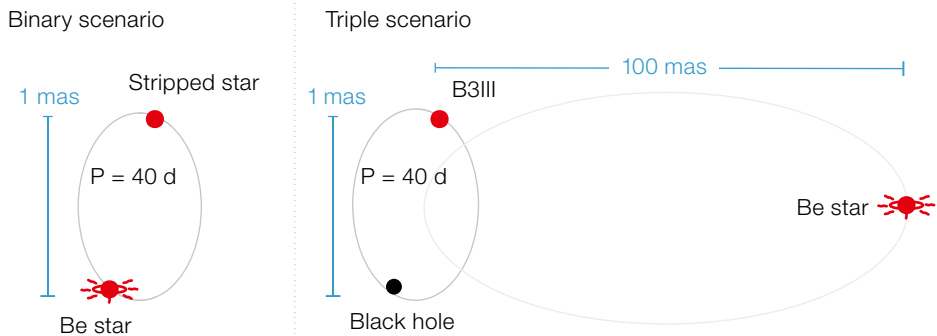

The broad absorption lines typical of a rapidly rotating star are more obvious in the high-quality spectra of HR 6819. In the 1980s, Dachs et al. (1981) and Slettebak (1982) had already noted the presence of both double-peaked emission lines due to a classical Be star, and narrow absorption features typical of a normal B-type giant in its spectrum. Maintz (2003) discovered that the narrow-lined star is on an orbit with a period of tens of days. With additional observations, Rivinius et al. (2020) showed that the star is moving on a 40-day period and reported that the absorption features of the Be star are consistent with its being stationary over a period of five years. Assuming a typical mass for the B-type star of ~ 5 $M_\odot$ and using the mass ratio obtained from the ratio of the RV semi-amplitudes would lead to an unphysical mass for the Be star if the two stars formed a binary. The Be star was thus interpreted as a third, outer component in a triple system, with the inner binary consisting of the B-type star and an unseen object of at least 4 $M_\odot$. As a normal star with this mass would have been visible in the spectra, this implied the presence of a stellar-mass black hole in the inner binary.

It was subsequently suggested that such a triple system would be unstable (Safarzadeh, Toonen & Loeb, 2020), and that the system could in fact be a quadruple (Mazeh & Faigler, 2020). However, the most robust challenge to the triple scenario came from three independent teams who re-analysed the optical spectra and found that the emission lines of the Be star do in fact show a subtle reflex motion with respect to the B-type star, as was found in LB-1 (Bodensteiner et al., 2020; El-Badry & Quataert, 2021; Gies & Wang, 2020). In this scenario, the B-type star and the Be star orbit around each other, removing the need for a third star or a black hole in the system. Figure 3 illustrates the different configurations of the binary and triple scenarios for HR 6819.

### Stripped stars in disguise

Both LB-1 and HR 6819 can be explained as binary systems of a B-type star and a classical Be star, rather than involving a stellar-mass black hole or a triple configuration. However, the main reason that the black hole scenarios were invoked in the first place is that the projected orbital velocities determined for the narrow absorption lines of the B-type stars are much larger than the ones estimated from the emission lines tracing the Be stars. These large velocity ratios indicate a large mass ratio, implying that the mass of the B-type star is not in the typical range of 5–6 $M_\odot$ but is instead a much lower ~ 0.5 $M_\odot$, which is unusually low for B-type stars. In this interpretation, HR 6819 and LB-1 are thus post-mass-transfer binary systems. The mass donor, the present B-type star, has become a low-mass stripped star caught in a puffed-up stage (see also Irrgang et al., 2020), while the mass gainer, the present Be star, accreted matter and angular momentum and was spun up to rapid rotation, but has a normal mass for its spectral type.

Stripped stars are the cores of stars that have lost part or most of their H-rich envelope either by strong stellar winds (only feasible for stars with masses ≥ 20 $M_\odot$; for example, Puls, Vink & Najarro, 2008) or, as is the case here, by mass transfer in interacting binary systems (for example, Podsiadlowski, Joss & Hsu, 1992). Once the mass transfer is complete, the now-exposed helium core has to adjust to the lack of a stellar envelope and is thus puffed-up. After a subsequent contraction phase lasting around a million years, a new equilibrium is attained. Figure 4 shows a possible evolutionary path of such a stripped star in comparison to a single star. The overlap of the two tracks, in particular in luminosity, with the observed parameters of HR 6819 shows why such stars could be mistaken for a normal main-sequence star.

In the new equilibrium phase, the stars are compact and hot (with surface temperatures between 50 000 and 100 000 K) and emit primarily in the far UV (Götberg, de Mink & Groh, 2017; Götberg et al., 2018). At the upper end of the mass regime, stripped stars can launch powerful winds that result in strong emission lines (Wolf-Rayet stars), making them easier to detect (Breysacher, Azzopardi & Testor, 1999; Crowther, 2007; Shenar et al., 2020a). At lower masses, these so-called subdwarf OB stars (Heber, 2009) are not expected to exhibit emission lines in the optical, making their detection difficult (Wellstein, Langer & Braun, 2001; Wang et al., 2021). The properties of stripped stars in an intermediate mass range between Wolf-Rayet and subdwarf OB stars are still debated as they have so far eluded detection.

Given the relatively short lifetime of the contraction phase — it amounts to around 1% of the main-sequence lifetime of the Be star — we expect to find stripped stars primarily in the subsequent evolutionary phase, which lasts about 10 times longer and during which the stars are hot and compact. The properties of the stars reported as stripped stars in HR 6819 and LB-1, however, match those of stars at the beginning of the contraction phase shortly after the mass transfer stopped (Shenar et al., 2020b; Irrgang et al., 2020; Bodensteiner et al., 2020, El-Badry & Quataert, 2021). In fact, the larger radius (and hence higher luminosity) and the lower temperature (shifting the peak of the emission into the optical wavelength range) of the puffed-up stripped star make it easier to detect systems spectroscopically in this phase, despite its shorter duration and the fact that the stripped stars only resemble B-type main-sequence stars at the beginning of the contraction phase. These stars have temperatures and luminosities similar to those of a main-sequence B-type star, but the envelope stripping has reduced their mass to around 0.5 $M_\odot$. An additional observational characteristic that helps to spot them is that stripped stars are expected to be slow rotators, manifesting as narrow and deep absorption lines, as observed in LB-1 and HR 6819.

### GRAVITY observations constraining the nature of HR 6819

Given the difficulties in measuring the orbital motion of the Be stars in LB-1 and HR 6819, both of the spectroscopic analyses presented above give plausible interpretations of the same observational dataset. Interferometric observations are the best way to obtain the definitive answer. The nearby ($d \sim 300$ pc) and bright ($V = 5.4$) HR 6819 system in particular is an ideal target for high-angular-resolution follow-up. The two proposed scenarios can be unambiguously distinguished by the spatial separation and motion of the two luminous sources: while they should only be 1–2 milliarcseconds apart and following a 40-day orbit according to the binary scenario, they should be significantly further apart and the Be star should be stationary on the month-long timescales of the observations according to the triple scenario with a black hole (see Figure 3).

Initially, speckle interferometry observations suggested the possible presence of a visible companion at around 120 milliarcseconds from the central source, favouring the triple scenario (Klement et al., 2021). Subsequently, higher-quality observations were executed using the Multi Unit Spectroscopic Explorer (MUSE) integral-field spectrograph at the Very Large Telescope (VLT) and the interferometric instrument GRAVITY at the VLT Interferometer (program ID 107.22R6, PI: Rivinius). While MUSE in its narrow-field mode covers the larger scales up to a few arcseconds, GRAVITY can resolve scales down to milliarcseconds.

As presented by Frost et al. (2022), the MUSE observations exclude the presence of a similarly bright source at around 120 milliarcseconds from the central source. The two GRAVITY epochs obtained approximately two weeks apart show two sources at a separation of ~ 1 milliarcsecond that switch position on the sky, as expected for a binary system on a 40-day orbit (see Figure 5). The two epochs thus unambiguously show that there are two luminous stars in a short-period binary, demonstrating that there is no black hole in the system.

Additional GRAVITY observations of HR 6819 (scheduled for April–September 2022, PI: Rivinius) will further allow the derivation of stellar parameters such as an accurate mass of the stripped star, which will be invaluable input for binary interaction models. Similarly, GRAVITY observations of LB-1 are scheduled before April 2022 (PI: Rivinius) and will hopefully pin down the nature of this system as well, although its larger distance makes it a more difficult target (if the binary scenario proves correct, the two luminous stars will most likely not be resolved).

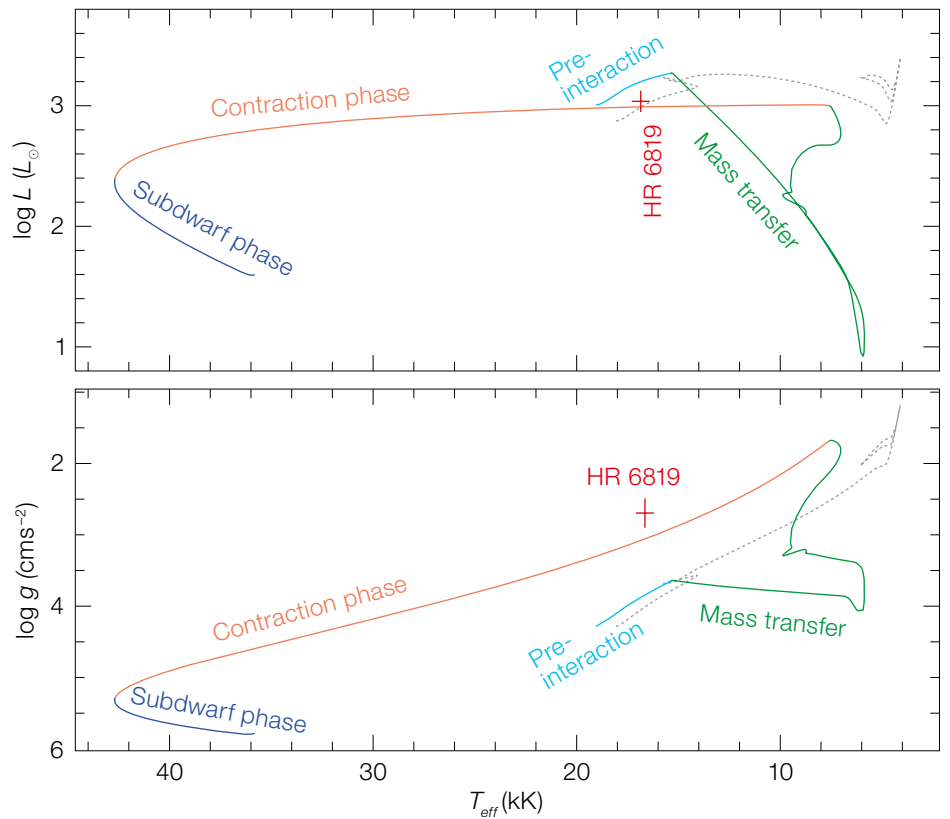

Figure 4. Possible evolutionary path of the stripped star in the HR 6819 system in a Hertzsprung-Russell (top panel) and a Kiel diagram (bottom panel), adapted from Bodensteiner et al. (2020). The different colours indicate different evolutionary phases and the red cross shows the observed properties of the stripped star in HR 6819. Note that the subdwarf phase lasts ~ 10 times longer than the contraction phase. Overplotted (in dotted grey) is the evolutionary track of a single star with an initial mass of 5.5 $M_\odot$. The Figure clearly shows that especially in temperature-luminosity space a contracting stripped star can be confused with a single main-sequence star, while the tracks are more separated in temperature–surface gravity space. A complete set of input files to reproduce the simulations underlying this figure can be found on Zenodo[2].

### The way forward for large surveys

The quest for the missing black holes continues. The LAMOST, SDSS, 4MOST and WEAVE large-scale surveys will undoubtedly detect many binaries with only one visible component and a high mass function, implying the presence of a massive unseen companion that could be a black hole.

The detection of stripped stars in HR 6819, LB-1 and possibly also NGC 1850 BH1 teaches us two important lessons. First, the mass of the initially discovered RV variable star is very important. Such masses are often estimated from a rough translation of spectral type into mass, but we now know that seemingly normal B-type stars can also be recently stripped stars with a much lower mass. We expect many more binaries that are actually post-interaction systems containing a stripped star in disguise, as well as their direct progenitors and successors, to be detected in large-scale surveys. Examples of the latter two categories are systems found during the interaction phase (such as recently proposed by El-Badry et al., 2022), and rapidly rotating stars with hot, compact companions that have already returned to equilibrium (for example, pre-white dwarfs or subdwarfs; Gies et al., 2020; Wang et al., 2021). Second, when searching for quiescent black holes via the movement of their stellar companions, one needs to look extremely carefully for signals of a rapidly rotating star in the spectra. The detection of broad emission lines can be a strong observational signature of a Be star, but can in some cases also be attributed to the presence of an accretion disc around a compact object. Broad absorption lines are unequivocal evidence for a rapidly rotating star (with or without a decretion disc) but are much more difficult to detect.

The way forward will have to be through an interdisciplinary approach, taking advantage of the wealth of available observational information. High-spectral-resolution, high-signal-to-noise observations, for example with the Ultraviolet and Visual Echelle Spectrograph (UVES; Dekker et al., 2000) and the Echelle SPectrograph for Rocky Exoplanet and Stable Spectroscopic Observations (ESPRESSO; Pepe et al., 2020) at the VLT, can help to identify the signatures of rapidly rotating companions as well as to obtain accurate parameters for the initially discovered RV variable stars. High-spatial-resolution spectroscopic follow-up, such as in the future with GRAVITY+ on the VLTI and the High Angular Resolution Monolithic Optical and Near-infrared Integral field spectrograph (HARMONI; Thatte et al., 2010) on ESO's Extremely Large Telescope (ELT), will be invaluable for distinguishing between scenarios with and without a black hole. Including multiwavelength data such as X-rays and radio may help to confirm the presence of a black hole accreting at very low rate. Long-term optical or near-infrared light curves can reveal low-amplitude orbital modulations that enable constraints on the orbital inclination to be set and may reveal a second luminous star even if it is not detected spectroscopically (Clavel et al., 2021; El-Badry, Rix & Heintz, 2021). Information from large-scale surveys, such as Gaia astrometry (and binary solutions in the future; Gaia Collaboration et al., 2016) or TESS photometry (Ricker et al., 2015) can also yield important pieces of the puzzle of understanding these newly detected systems. By employing these methods, we can not only better understand rare, post-interaction sources and thus binary evolution but also hone the search for black holes. After all, the presence of an intrinsically dark and quiet object can be only indirectly proven by rejecting all other possibilities.

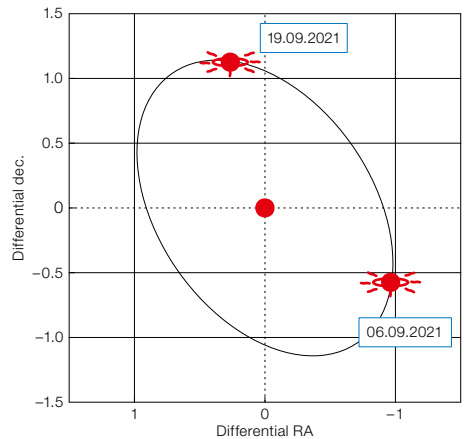

Figure 5. Relative astrometric orbit (black) of HR 6819 presented in Frost et al. (2022). The B-type star is fixed at coordinate (0,0) and the position of the Be star measured at two different epochs is indicated. It is obtained by simultaneously fitting the measured RVs of the B-type star and the astrometry obtained from interferometric observations with GRAVITY.

### Acknowledgements

The authors would like to thank everyone involved in the HR 6819 project for their input and fruitful discussions. JB, MH and MAM are supported by an ESO Fellowship. DMB acknowledges a senior postdoctoral fellowship from the Research Foundation Flanders (FWO) with grant agreement no. 1286521N. PM acknowledges support from the FWO junior postdoctoral fellowship no. 12ZY520N. LM thanks the European Space Agency and the Belgian Federal Science Policy Office for their support in the framework of the PRODEX Programme. TS acknowledges support from the European Union's Horizon 2020 under the Marie Skłodowska-Curie grant agreement no. 101024605.

Links

[1] GRAVITY+ White Paper: https://www.mpe.mpg.de/7480772/GRAVITYplus_WhitePaper.pdf

[2] Data products and input files necessary to reproduce the simulations shown in Figure 4: https://doi.org/10.5281/zenodo.5875376

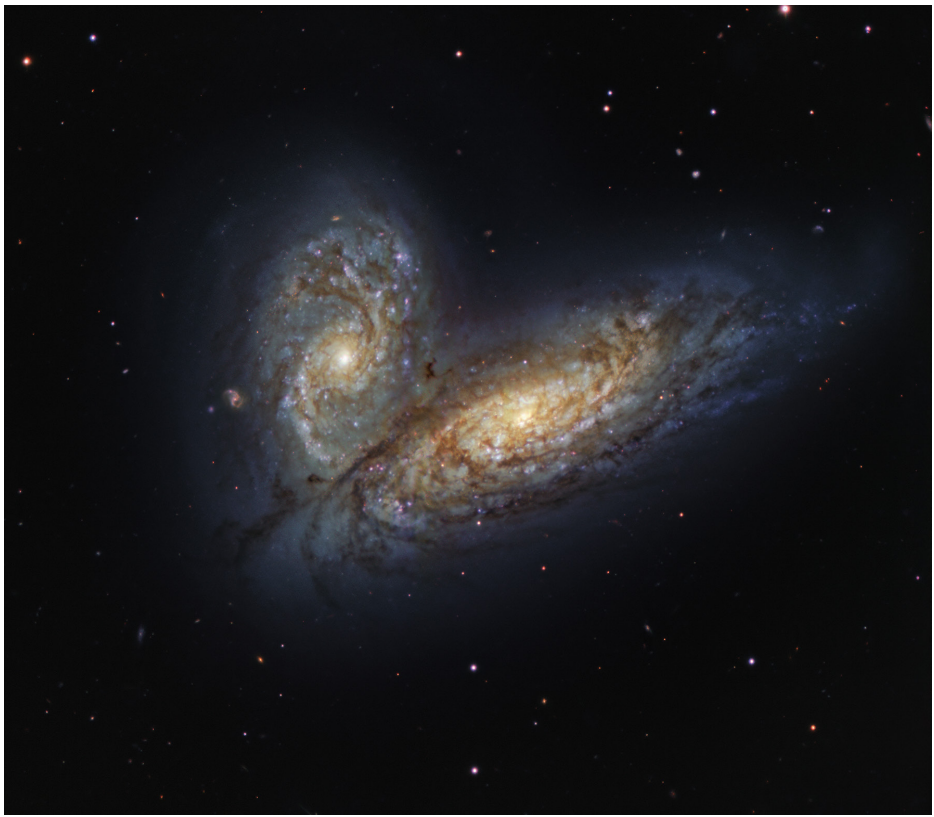

Around 60 million light-years away, in the constellation Virgo, the two galaxies NGC4567 and NGC4568, nicknamed the Butterfly Galaxies owing to their wing-like structure, are beginning to collide and merge into each other. This is depicted in this picture captured by the FOcal Reducer and low dispersion Spectrograph 2 (FORS2) instrument, which is mounted on ESO's Very Large Telescope (VLT) at Paranal Observatory in the Chilean Andes.